\documentclass[prb,twocolumn,showpacs,amsmath,amssymb,superscriptaddress]{revtex4-1}
\usepackage{graphicx}% Include figure files
\usepackage{dcolumn}% Align table columns on decimal point
\usepackage{bm}
\usepackage{natbib}
\usepackage{amssymb}
\usepackage{amsmath}
\usepackage{dcolumn}
\usepackage[usenames,dvipsnames]{xcolor}

\begin{document}

\title{Fast Free Energy Calculations for Unstable High-Temperature Phases}
\author{Nikolas Antolin}
\affiliation{Dept.\ of Materials Science and Engineering, The Ohio State University, Columbus, OH 43210}
\author{Oscar D. Restrepo}
\affiliation{Dept.\ of Materials Science and Engineering, The Ohio State University, Columbus, OH 43210}
\author{Wolfgang Windl}
\affiliation{Dept.\ of Materials Science and Engineering, The Ohio State University, Columbus, OH 43210}

\date{\today}

\begin{abstract}
We present a fast and accurate method to calculate vibrational properties for mechanically unstable high temperature phases that suffer from imaginary frequencies at zero temperature. The method is based on standard finite-difference calculations with optimized large displacements and is significantly more efficient than other methods. We demonstrate its application for calculation of phonon dispersion relations, free energies, phase transition temperatures, and vacancy formation energies for body-centered cubic high-temperature phases of Ti, Zr, and Hf.
\end{abstract}
\maketitle

\begin{center}
{\textbf{{\normalsize I. Introduction}}}
\end{center}

Hexagonal close packed (hcp) to body-centered cubic (bcc) phase transitions are characteristic of a number of transition and rare earth metals, including titanium, zirconium, and hafnium. Although room-temperature applications of these materials are limited by the expense of processing them, their high-temperature properties make them ideal candidates as the basis of high-temperature alloys. However, only limited experimental data \cite{4,5,6} have been collected on the bcc high-temperature phase of these systems, and they are inaccessible to traditional zero-temperature ab-initio calculations due to their mechanical instability.\cite{WW1} Although the recently proposed self-consistent ab-initio lattice dynamics (SCAILD) \cite{1} and molecular dynamics (MD) based \cite{3} approaches  have proved successful in calculating phonon dispersions for these materials, convergence issues with free energy and phonon slopes in the long-wavelength limits prevent accurate calculation \cite{2} of other phase properties relevant to high-temperature applications (e.g. elastic constants and formation energies of vacancies). As a result, the computational cost of these advanced calculation procedures sets limits to their applicability, and a supplemental calculation method is necessary for complete phase analysis. Thus, there is significant motivation to develop a simplified method of dealing with mechanically unstable systems of potentially large size that quickly produces reasonable ab-initio results.

Current ab-initio methods to deal with mechanical instability in these systems have focused on a simulated MD approach by displacing atoms around their equilibrium positions and using the resulting structures to calculate dynamical matrices for phonon calculations. The SCAILD method proposed by Souvatzis et al.\ \cite{1} is limited by its high symmetry requirement for reasonable calculation convergence, and is thus inefficient at producing calculations beyond phonon dispersions for perfect structures.\cite{2} A second calculation method proposed by Hellman et al. utilizes an MD approach to extract dynamical matrices, but is computationally intensive due to the large number of MD steps it necessitates for convergence.\cite{3} Due to their computational demands, neither method has been used to date to calculate the energetics of defects in mechanically unstable systems.

In this letter, we propose a method to calculate accurate thermodynamic properties of mechanically unstable high-temperature phases using ab-initio theory. This method utilizes large-displacement phonon calculations with a displacement determined by analysis of the phonon amplitude-energy curve. Because these calculations are computationally very economical, more complex calculations of defect energies, alloying effects and transition temperatures are possible than using previous ab-initio methods. Furthermore, we find for the high-temperature bcc phases that we have studied that the agreement of phonon dispersions and phase transition temperatures with experimental values is at least as good  as with previously published ab-initio methods.

In high-temperature bcc phases, where the mechanical instability occurs as a result of decreased energy when a ($\varepsilon$,$-\varepsilon$,0,0,0,0) strain is applied to the calculation cell \cite{WW1}, the energy-strain response in this direction can be approximated by a quartic function with negative curvature at zero strain. The corresponding elastic constant $c'$, given by that curvature, is thus negative. Due to the relationship between the slopes of the (acoustic) phonon branches and the elastic constants, especially for the transverse branch in  $[1{\bar 1}0]$ direction which is given by $\left(C_{11}-C_{12}\right)/2$, the negative elastic constant corresponds to imaginary frequencies in the phonon dispersion, which is determined from the basic assumption that the energy vs.\ atomic displacement curves have positive curvature (minimum at the equilibrium position) and are parabolic (harmonic). Phonon frequencies and eigenvectors are determined by the curvature of the energy vs.\ atomic displacement curve at the equilibrium position and can be determined by density-functional perturbation theory \cite{WW2} or small finite displacements.\cite{WW3}

The appearance of imaginary frequencies does not only interfere with determining phonons per se, but also with free-energy based calculations, which for solids are easiest obtained within the quasiharmonic approximation. There, the free energy contribution to the total energy is calculated from the phonon density of states. \cite{WW2} A small-displacement phonon calculation of the free energy contribution to mechanically unstable structures is thus flawed through its inclusion of nonphysical frequencies. 

The fundamental question at this point is why the high-temperature phases, despite being mechanically unstable, become stable at sufficiently high temperatures. As has been discussed previously, this is due to the fact that the vibrational amplitudes of the atoms increase with temperature. This has two consequences: {firstly}, an atom {no longer} moves in the potential landscape of the other atoms at or close to their equilibrium {lattice} position{s}, but displaced from {them} (phonon-phonon coupling). Secondly, the atom explores a larger part of the energy landscape through its higher {vibrational} amplitude (phonon anharmonicity). If the second effect dominated the first, a calculation that includes these anharmonic effects could be based on an improved displacement strategy around the equilibrium positions of the atoms. This would avoid the major computational overhead in the methods proposed to date \cite{1,3}, which consist of relating displacements for the other atoms in the system to the atom displaced for a phonon calculation. Indeed, such a strategy may be successful and not hampered by the instability in direct vicinity of the equilibrium position considering that an oscillating atom has its largest velocity when it passes through the equilibrium site, while its velocity decreases to zero at the turnaround points, where it thus spends the majority of its time. Therefore, we propose to use atomic displacements for the phonon calculations large enough to probe the curvature of the energy landscape far away from the equilibrium position.

\begin{figure}
\includegraphics[width=1.0 \columnwidth, angle=0, scale=1]{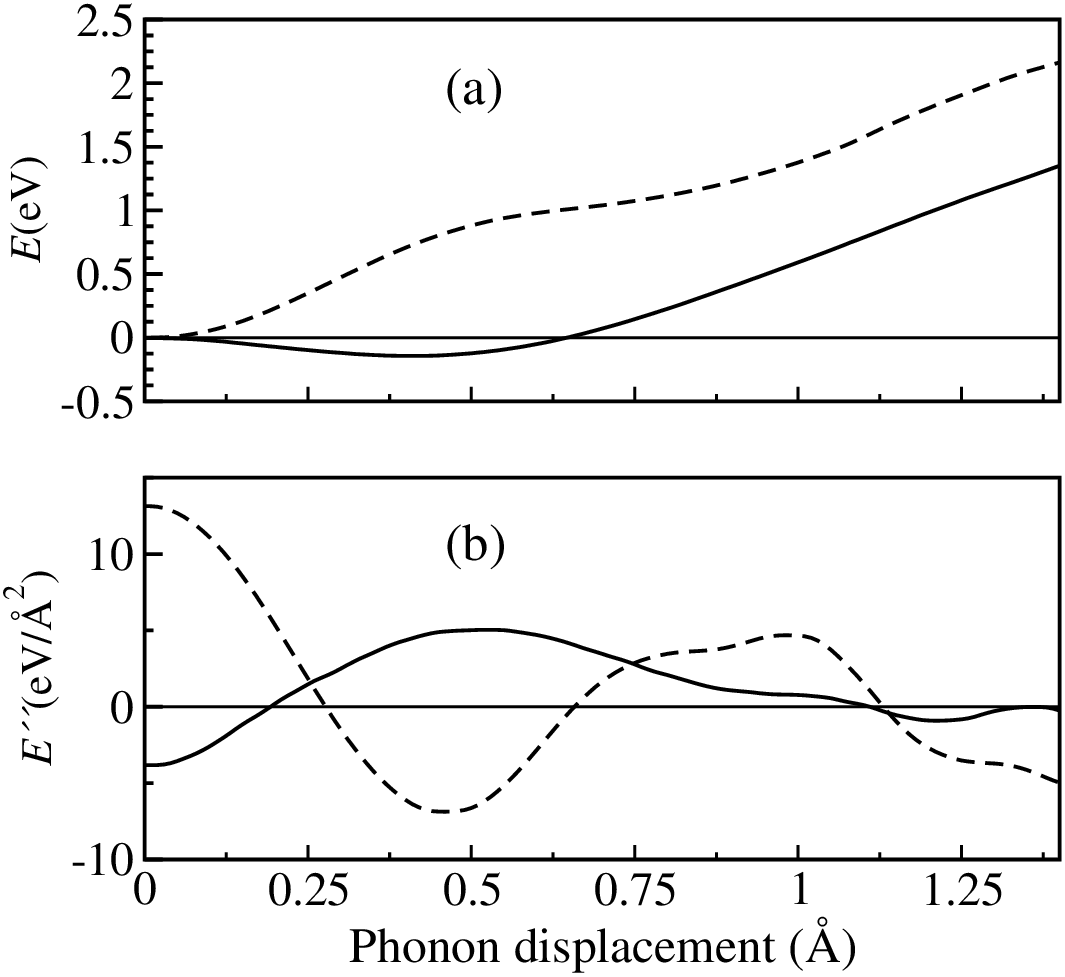} 
\caption{The (a) energy-vs.-phonon amplitude curve and its (b) second derivative for an $N$-point frozen phonon cell \cite{8} displacement of the bcc cell of Hf (solid) and Mo (dashed).}
\label{Fig1}
\end{figure}

Indeed, at large displacements, the quartic term of the energy-strain function eclipses the negative quadratic term, leading to an approximately constant positive curvature at sufficient distance (e.g. at $\sim$0.5 {\AA} in Fig.~\ref{Fig1}). Since a constant curvature corresponds to harmonic motion and enables the phonon concept, the shape of the curve can thus be approximated by only using a quadratic term with little loss of detail. This suggests that a phonon calculation of the free energy contribution using a large displacement in the quartic-dominated regime could be used with the quasi-harmonic approximation to provide the free energy contribution to the total structure energy, and that the properties of a structure thus stabilized could be calculated using ab-initio methods. By comparing the results of a large-displacement calculation to experiment, one then can also probe in a simple way if the dominating anharmonic effect comes from the large atomic displacements (with lesser effect from what the surrounding atoms do), or vice versa.

%%%%%%%%%%%%%%%%%%%%%%%%%%%%%%%%%%%%%%%%%%%%%%%%%%%%%%%%%%%%%%%%%
%%%%%%%%%%%%%%%%%%%%%%%%%%%%%%%%%%%%%%%%%%%%%%%%%%%%%%%%%%%%%%%%%

\begin{center}
{\textbf{{\normalsize II. Methodology}}}
\end{center}

To calculate the approximate large displacement necessary to force the phonon calculation into the quartic-dominated regime, VASP \cite{7} was used on a frozen-phonon type supercell for $N$-point phonons \cite{8} to calculate the internal energy of the structure as a function of phonon displacement.\cite{WW5} The $N$-point phonon was chosen since the transverse mode at the $N$-point is the primary marker of mechanical instability in bcc transition metals.\cite{7} For these calculations, we used projector-augmented wave potentials \cite{PAW1,PAW2} with a 10$\times$10$\times$10 Monkhorst-Pack k-point mesh. Calculations were performed for bcc titanium, zirconium and hafnium, as well as several mechanically stable bcc metals.

Figure~\ref{Fig1}(a) shows the calculated internal energy as a function of the $N$-point phonon amplitude in the chosen displacement direction for the Hf and Mo  systems. To determine the appropriate phonon displacement for large-displacement calculations, second numerical derivatives of this internal energy data are calculated using a central-difference method, shown in Figure 1(b). In the case of Hf, these calculations give an indication of the point at which the amplitude-energy curve enters the quasi-parabolic high-temperature regime (in this case between $\sim$0.4 and 0.7 {\AA}, where the second derivative becomes approximately constant), allowing harmonic-type phonon calculations and use of the quasi-harmonic approximation to determine the system's free energy. By contrast, the second derivative for Mo is maximally positive at values around zero displacement, which therefore defines its quasiharmonic regime. For automatic detection of an appropriate displacement value, we found that the maximum value of the second derivative identifies well the necessary displacement to calculate a realistic phonon dispersion. This interpretation matches well with phonon approaches for stable materials, which exhibit maximum second derivative values at minimal displacements. Alternatively, one can select the displacements just large enough that all the imaginary frequencies from the phonon dispersions are eliminated. Typically, the last frequencies to turn real are located not at the $N$-point, but close to the $\Gamma$-point.%, and have minimal effect on the integrated phonon DOS.

%%%%%%%%%%%%%%%%%%%%%%%%%%%%%%%%%%%%%%%%%%%%%%%%%%%%%%%%%%%%%%%%%%%%
\begin{table}[htdp]
\caption{Calculated vs. experimental values (in parentheses) [\protect\onlinecite{9,11,4,5,6,CRC}] of phase transition temperature $T_{\rm trans}$, bulk modulus $K$, and lattice parameter $a$ for the bcc high-temperature phases of Ti, Zr, and Hf. Also given are large-displacement values determined from maximum values of the second derivative of $N$-point frozen-phonon cells ($u_{\rm 2d}$) and from a no-imaginary-frequency condition ($u_{\rm nI}$) used for these calculations. }
\begin{center}
\begin{tabular}{cccccc}
\hline
\hline
  &  $u_{\rm 2d}$ ({\AA}) & $u_{\rm nI}$ ({\AA}) & $T_{\rm trans}$ (K) & $a$ ({\AA}) & $B$ (GPa) \\
\hline
Ti & 0.60 & 0.88 & 1200 (1155) & 3.276 (3.33)\phantom{0} & 87.1 (87.7) \\
Zr & 0.57 & 0.54 & 1050 (1136) & 3.557 (3.551) & 94.8 (96.7) \\
Hf & 0.63 & 0.68 & 2100 (2016) & 3.521 (3.625) & 119 (112.3) \\
\hline
\hline
\end{tabular}
\end{center}
\label{Table1}
\end{table}
%%%%%%%%%%%%%%%%%%%%%%%%%%%%%%%%%%%%%%%%%%%%%%%%%%%%%%%%%%%%%%%%%%%%

Once the necessary displacement magnitude was determined for each system by eliminating all the imaginary frequencies from the phonon dispersion, force constants and dynamical matrices were determined from finite-displacement VASP calculations within 4$\times$4$\times$4 supercells of the bcc primitive cells using the PHON code.\cite{WW3} The calculations were performed with 6$\times$6$\times$6 Monkhorst-Pack k-point meshes for a range of lattice parameters. Using the quasi-harmonic approximation in PHON \cite{WW3}, these calculations were used to determine the vibrational contribution to the system's free energy over a temperature range that included the experimental transition temperature. For all materials considered, an additional calculation of the phonon density of states was conducted on relaxed supercells composed of 3$\times$3$\times$2 hexagonal unit cells with a $\Gamma$-centered 6$\times$6$\times$9 Monkhorst-Pack k-point mesh using a 0.03 {\AA} phonon amplitude. The lattice parameter was then determined by minimizing the free energy with respect to volume. The minimum  of the free energy vs. volume curve at each temperature was then used to determine the phase transition temperature. The bulk modulus was calculated using a Birch-Murnaghan fit for temperatures across the literature range of phase stability and plotted to determine its temperature dependence.

To calculate the free energy of formation of a vacancy in the metallic bcc system, $F_f(T) = F_V(T)-\frac{N-1}{N} F_p(T)$, the free energy as a function of temperature for 3$\times$3$\times$3 54-atom bcc super cells with ($V$) and without ($p$) vacancies was calculated using a 6$\times$6$\times$6 Monkhorst-Pack k-point mesh and the large-displacement values from the perfect cells. For this demonstration here, the vacancy structures were not relaxed.  
%The free energy of vacancy formation at the transition temperature, an intermediate temperature, and at the melting point was then calculated from the %resulting free energy and that of a perfect cell containing the same number of atoms.

Table~\ref{Table1} shows the calculated transition temperature, lattice parameter, and bulk modulus obtained using the large displacement phonon approach wherein no imaginary frequencies were present in the phonon dispersion, as well as experimental values. In addition, the displacements are displayed at which no imaginary frequencies appear in the phonon dispersion, and at which the second derivative of the phonon amplitude-energy curve was maximal.

%%%%%%%%%%%%%%%%%%%%%%%%%%%%%%%%%%%%%%%%%%%%%%%%%%%%%%%%%%%%%%%%%
\begin{figure}
\includegraphics[width=1.0 \columnwidth, angle=0, scale=1]{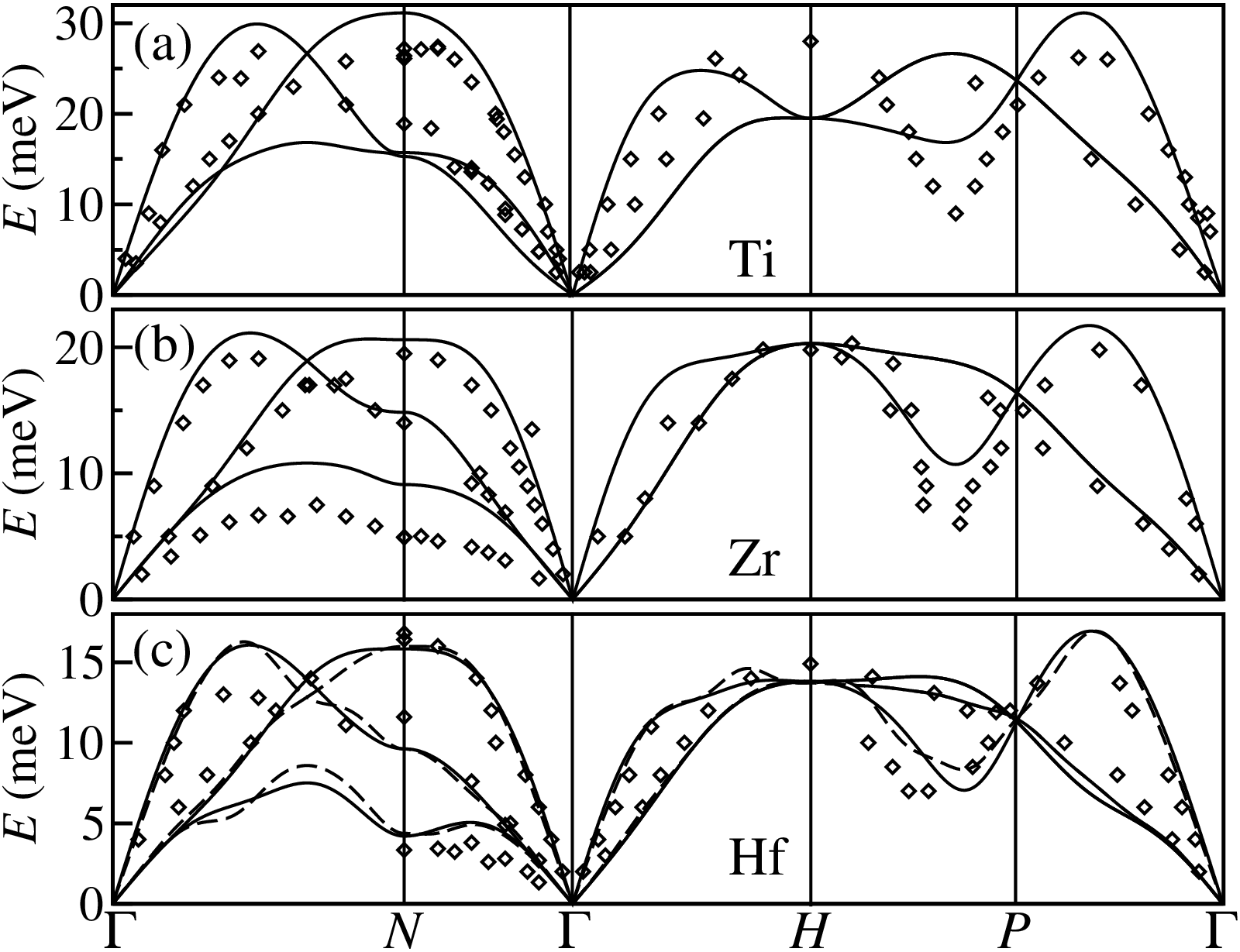} 
\caption{Large-displacement calculated phonon dispersions for (a) Ti, (b) Zr, and (c) Hf vs.\ experimental phonon energies from [\protect\onlinecite{4,5,6}]. Solid lines are calculated with 4$\times$4$\times$4 phonon supercells, the dashed line for Hf with 8$\times$8$\times$8.}
\label{Fig2}
\end{figure}
%%%%%%%%%%%%%%%%%%%%%%%%%%%%%%%%%%%%%%%%%%%%%%%%%%%%%%%%%%%%%%%%%

\begin{center}
{\textbf{{\normalsize III. Results}}}
\end{center}

For bcc titanium, the first displacement at which no imaginary frequencies were observed in the phonon dispersion was 0.88~{\AA}; this value did not correspond to the absolute maximum of the second derivative, which occurred at approximately 0.6~{\AA}, but did correspond to a second local maximum of the second derivative which occurred at approximately 0.85~{\AA}. The transition temperature was calculated to be 1200 K, which is within 4\% of the experimental value of 1155 K \cite{9}, and closer to experiment than that calculated using the SCAILD method, 1107 K \cite{2}. The lattice parameter was calculated to be 3.234~{\AA} at 0 K and 3.276~{\AA} at the calculated transition temperature, while the experimental lattice parameter at this temperature was 3.33~{\AA}. The coefficient of thermal expansion was calculated to be $6.7\times10^{-6}$ $\textnormal{K}^{-1}$, compared to an experimental coefficient of thermal expansion for the BCC phase of $8.6\times10^{-6}$ $\textnormal{K}^{-1}$.\cite{NA1} The calculated phonon dispersion using the 0.88~{\AA} phonon amplitude is displayed in Figure~\ref{Fig2}(a), with experimental data points from Ref. \onlinecite{4}.

For bcc zirconium, the first displacement at which no imaginary frequencies were observed in the phonon dispersion was 0.54~{\AA}; this value agreed well with the calculated maximum of the second derivative, which was approximately 0.57~{\AA}. The transition temperature was calculated to be 1050 K, which is 8\% lower than the experimental value of 1136 K \cite{9}, whereas the transition temperature calculated using the SCAILD method, 916 K \cite{2}, is more than 19\% too low. The lattice parameter was calculated to be 3.557~{\AA} at the calculated transition temperature and 3.562~{\AA} at 1150 K, while the experimental lattice parameter near the transition temperature was 3.551~{\AA}. The coefficient of thermal expansion was calculated to be $1.34\times10^{-5}$ $\textnormal{K}^{-1}$ (no corresponding experimental measurements were found). The calculated phonon dispersion using the 0.54~{\AA} phonon amplitude is displayed in Figure~\ref{Fig2}(b), in comparison to experimental data points.\cite{5}

For bcc hafnium, the first displacement at which no imaginary frequencies were observed in the phonon dispersion was 0.68~{\AA}; this value agreed well with the calculated maximum of the second derivative, which was approximately 0.63~{\AA}. The transition temperature was calculated to be 2100 K, which is 4\% higher than the experimental value of 2016 K \cite{11}, while the transition temperature calculated using the SCAILD method, 1660 K \cite{2}, is 18\% too low. The lattice parameter at 2400~K was calculated to be 3.521~{\AA}, which is 3\% smaller than the experimental value of 3.625~{\AA} at that temperature. The coefficient of thermal expansion was calculated to be $7.9\times10^{-6}$ $\textnormal{K}^{-1}$, compared to an experimental coefficient of thermal expansion for the BCC phase of $11\times10^{-6}$ $\textnormal{K}^{-1}$.\cite{NA2} The calculated phonon dispersion using the 0.68~{\AA} phonon amplitude is displayed in Figure~\ref{Fig2}$({\rm c})$, with experimental data points \cite{6}. Additional phonon calculations were performed on the Hf bcc system using larger supercells to determine whether increases in cell size would improve the agreement of results with experimental values. Within the feasible limits of computational resources available for this study, no significantly better agreement was observed from increasing the calculation cell size past a supercell composed of 4$\times$4$\times$4 bcc primitive cells. The results of expanding the supercell size from 4$\times$4$\times$4 to 8$\times$8$\times$8 bcc primitive cells is shown in Fig.~\ref{Fig2}(c).

%{\bf COMPARISON OF OUR DISPLACEMENTS TO SCAILD DISPLACEMENTS!?}

\begin{figure}
\includegraphics[width=1.0 \columnwidth, angle=0, scale=1]{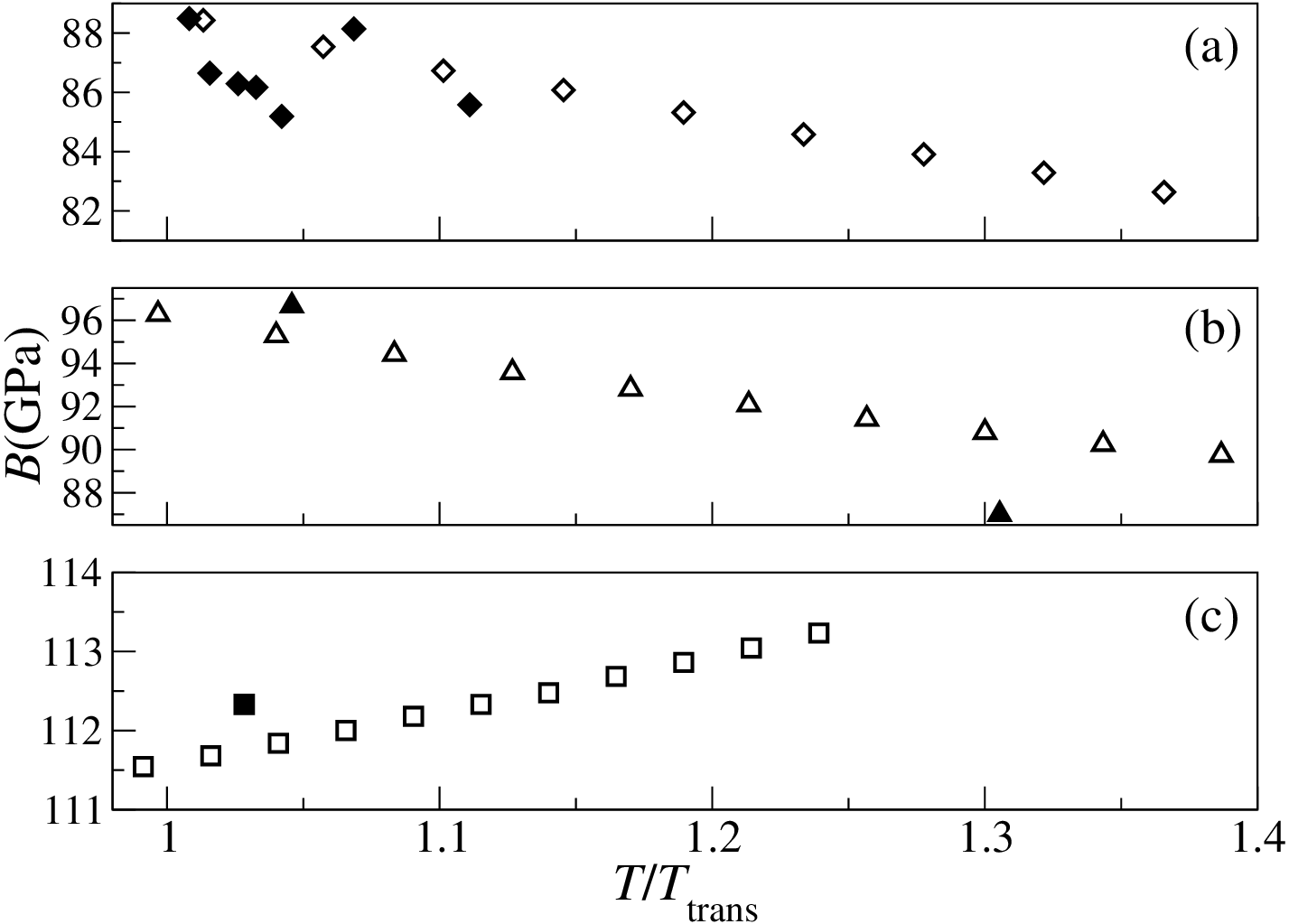} 
\caption{Calculated temperature dependence of bulk modulus for (a) Ti, (b) Zr, and (c) Hf, normalized to the transition temperature of each metal, with corresponding experimental values from [\onlinecite{NA1,5,6}].}
\label{Fig3}
\end{figure}

The temperature dependence of bulk moduli for Ti, Zr, and Hf was calculated using the large-displacement method with Birch-Murnaghan fitting across a temperature range as previously described. The three temperature dependencies, displayed as normalized to each material's transition temperature in Figure~\ref{Fig3}, exhibit nearly perfect quadratic relationships. Also plotted in Figure~\ref{Fig3} are experimental data for the bulk moduli of the three materials, shown as solid points. Calculated values for Ti and Zr exhibit the typically expected decrease in bulk modulus with increased temperature. The experimental data for Ti (Fig.~\ref{Fig3}(a)) show a large scattering and indicate significant error bars for the measured values, which are not reported in the experimental source [\onlinecite{NA1}]. Within these assumed error bars, our temperature dependence is in agreement for Ti and Zr, where values have been measured for more than one temperature. Calculated values for Hf indicate an anomalous increase in bulk modulus with increased temperature and show good agreement with the one experimental value available [\onlinecite{6}] for a temperature of 2073K. While it is possible that this represents a calculation artifact or results from the Birch-Murnaghan fits used, if physically valid it would indicate an internal structural change that may be related to the mechanism of the HCP-BCC phase transition.  

%More info on this needed, perhaps B v. T behavior of Mo or W

The free energy of formation for unrelaxed vacancies in bcc Ti, Zr, and Hf were calculated using the large-displacement method with the displacements previously described. By fitting {$F_f(T) = E_f-TS_f$} \cite{Nathan} in the bcc stability range, the energy $E_f$ and entropy $S_f$ of formation of the vacancy were determined. In Ti, the vacancy energy of formation $E_f$ was 2.05~eV, and the entropy of formation $S_f$ was $8.15k_B$. In Zr, the vacancy energy of formation $E_f$ was 2.06~eV, and the entropy of formation $S_f$ was $2.76k_B$. In Hf, we found $E_f =  2.20~eV$ and $S_f = 0.89k_B$. The values for the formation energy determined in this way are smaller by 0.9\%, 0.6\%, and 5.9\% than the zero-temperature formation energies of 2.07, 2.08, and 2.34 eV for Ti, Zr, and Hf respectively. While the agreement between fit and zero-temperature formation energy for Ti and Zr is similar to more harmonic bcc metals, where we calculate a decrease of 0.2\%, 0.4\%, and 1.0\% for the examples of W, Mo, and Cr, respectively,\cite{AntolinTBP} Hf shows a strong discrepancy coming from the fact that its phonon density of states shows a strong decrease in the high-frequency regime upon introduction of a vacancy which is not observed for Ti and Zr. Additionally, it would not be possible to calculate the entropy of vacancy formation without a reliable temperature dependence of free energy derived from a converged phonon density of states, as calculated using this method. We find that the entropy of formation scales inversely with the melting temperature of the three metals, which is sensible, since the melting temperature is a measure for the influence of the lattice vibrations on the stability of the lattice  the higher the melting temperature, the smaller the effect of lattice vibrations on the order of the atoms with respect to their equilibrium position, and thus the smaller the influence of vibrational disorder on the formation of vacancies. In order to reliably calculate the relative magnitude of these contributions for anharmonic structures at high temperatures, full free-energy calculations are a necessity, and our results show that they are well within the capabilities of the method proposed here.

%Thus, in order to calculate reliable defect energetics, zero-temperature calculations that are well trusted for harmonic materials need to be replaced by full free-energy calculations for such anharmonic metals{, which are well feasible with the method proposed here.}

%The strong anharmonicity in Zr and Hf can also be seen when allowing for a linear temperature dependence of the entropy during the free-energy fit in the %form $S(T)=S_0+S_1T$. While this increases the fitted formation energy for Ti and Hf by 3.0\% and 3.1\% in comparison to the fit with constant entropy, %much smaller decreases of $-0.2$\%, $-0.6$\%, and $-0.1$\% are found for Mo, Cr, and W. Even more dramatic, the constant entropy term  $S_0$  in the %fit with temperature-dependent entropy is 20\% and 80\% larger for Ti and Hf than the values fitted with constant entropy, whereas very small decreases %of $-0.4$\%, $-1.2$\%, and $-1.9$\% are found for W, Mo, and Cr.

%%%%%%%%%%%%%%%%%%%%%%%%%%%%%%%%%%%%%%%%%%%%%%%%%%%%%%%%%%%%%%%%%
%%%%%%%%%%%%%%%%%%%%%%%%%%%%%%%%%%%%%%%%%%%%%%%%%%%%%%%%%%%%%%%%%

\begin{center}
{\textbf{{\normalsize IV. Conclusions}}}
\end{center}

In this paper, the large-displacement phonon approach has been shown to produce results in good agreement with experimental values for a number of high-temperature metallic systems. However, it cannot be expected to be applicable to all systems in the same way presented here, as the causes underlying mechanical instability in these systems may differ, although preliminary calculations for other systems up to now look promising. Our methodology can only be applied to those systems in which a parabolic regime can be reached by increased phonon amplitude,
% in this high-symmetry phonon direction, 
as determined in the displacement calculation steps. For these systems, it produces results that have excellent agreement with experimental values at a relatively low computational cost. For the other systems, large-displacement calculations can help to easily show that phonon-phonon interactions dominate the effect of a locally anharmonic potential.

%than other methods for correcting anharmonic phonon behavior. As a result, it allows more complex calculations to be performed, including defect calculations, that were previously beyond the reach of ab-initio computational methods.

{In summary, the primary advantages in the large-displacement method for calculating the energetics of mechanically unstable high temperature phases and defects are its speed and simplicity. Due to the computational demands of both the SCAILD \cite{1} and MD \cite{3} approaches, the cell volume is at best estimated from an extrapolation from a 0 K DFT relaxation, which can lead to large errors as shown here. In contrast, the large-displacement method provides a fast self-consistent approach that allows phonon calculations on a number of different calculation cell volumes simultaneously, allowing for true incorporation of temperature effects due to lattice expansion and atomic vibration to yield more accurate values for thermodynamically determined quantities such as phase transition temperatures. The incorporation of these effects via the large-displacement method seems to allow for results in the same range of accuracy as those obtained for mechanically stable phases.} {Additionally, the significantly reduced computational cost allows for more complex calculations to be performed, including defect calculations, that were previously beyond the reach of ab-initio methods.} 

Funding for this work is acknowledged from the Center for Emergent Materials at the Ohio State University, an NSF MRSEC (Award Number DMR-0820414), and partial funding from the Air Force Office of Scientific Research under Contract Number FA9550-09-1-0251.


\begin{references}

\bibitem{4} W.\ Petry {\it et al.}, Phys.\ Rev.\ B {\bf 43}, 10{\thinspace}933 (1991).

\bibitem{5} A.\ Heiming {\it et al.}, Phys.\ Rev.\ B {\bf 43}, 10{\thinspace}948 (1991).

\bibitem{6}J.\ Trampenau {\it et al.}, Phys.\ Rev.\ B {\bf 43}, 10{\thinspace}963 (1991).

\bibitem{WW1} M.\ F.-X.\ Wagner and W.\ Windl, Acta Mater.\ {\bf 56}, 6232 (2008).

\bibitem{1} P.\ Souvatzis, O.\ Eriksson, M.\ I.\ Katsnelson, and S.\ P.\ Rudin, Comput.\ Mater.\ Sci.\ {\bf 44}, 888 (2009).

\bibitem{3} O.\ Hellman, I.\ A.\ Abrikosov, and S.\ I.\ Simak, Phys.\ Rev.\ B {\bf 84}, 180{\thinspace}301 (2011). 

\bibitem{2} P.\ Souvatzis, S.\ Arapan, O.\ Eriksson, and M.\ Kastnelson, arXiv:1102.2139v2 [cond-mat.stat-mech].

\bibitem{WW2} S.\ Baroni, S.\ de Gironcoli, A.\ Dal Corso, and P.\ Giannozzi, Rev.\ Mod.\ Phys.\ {\bf 73}, 515 (2001).

\bibitem{WW3} D. Alf\'e, Comput.\ Phys.\ Commun.\ {\bf 180}, 2622 (2009).

\bibitem{8} M. Mehl {\it et al.}, Phonons in bcc Crystals -- Point 4 ($N$), http://cst-www.nrl.navy.mil/users/mehl/phonons/ bcc/point04/, accessed on 10/17/11.

\bibitem{7} G.\ Kresse and J.\ Hafner, Phys. Rev. B {\bf 47}, 558 (1993); {\bf 49}, 14251 (1994).

\bibitem{WW5} VASP and many other ab-initio codes by default only allow calculations for small displacements. To avoid the code overriding the input of a large displacement, it needs a simple modification. 

\bibitem{PAW1} P.\ E.\ Bl\"ochl, Phys.\ Rev.\ B {\bf 50}, 17{\thinspace}953 (1994).

\bibitem{PAW2} G.\ Kresse and D.\ Joubert, Phys.\ Rev.\ B {\bf 59}, 1758 (1999).

\bibitem{9}J.\ L.\ Murray, in Binary Alloy Phase Diagrams, ed.\ T.\ B.\ Massalski (ASM International, Materials Park, OH, 1990) 2nd ed., Vol.\ 3, p.\ 3502.

\bibitem{11}J.\ P.\ Abriata, J.\ C.\ Bolcich, and H.\ A.\ Peretti, in Binary Alloy Phase Diagrams, ed.\ T.\ B.\ Massalski (ASM International, Materials Park, OH, 1990) 2nd ed., Vol.\ 3, p.\ 2128.

\bibitem{CRC} CRC Handbook of Chemistry and Physics, ed.\ D.\ R.\ Lide (CRC Press, Boca Raton, FL, 2006-7), 87th ed.

\bibitem{Nathan} N.\ G.\ Stoddard, P.\ Pichler, G.\ Duscher, and W.\ Windl, Phys.\ Rev.\ Lett.\ {\bf 95}, 025901 (2005); W.\ Windl, ECS Trans.\ {\bf 16}, 89 (2008).

\bibitem{NA1} H.\ Ogi, S.\ Kai, H.\ Ledbetter, R.\ Tarumi, and M.\ Hirao, Acta Mater.\ {\bf 52}, 2075 (2004).

\bibitem{NA2} P.\-F.\ Paradis, T.\ Ishikawa, and S.\ Yoda, International Journal of Thermophysics {\bf 24}, 1{\thinspace}239 (2003).

\bibitem{AntolinTBP} N.\ Antolin and W.\ Windl (to be published).

%\bibitem{Korhonen95}  T. Korhonen, M. J. Puska, and R. M. Nieminen, Phys.\ Rev.\ B {\bf 51}, 9526 (1995). 

\end{references}
\end{document}